\UseRawInputEncoding
\documentclass[UTF8]{revtex4}
\usepackage{epsfig}
\usepackage{psfrag}
\usepackage{amsfonts}
\usepackage{graphicx}
\usepackage{dcolumn}
\usepackage{bm}
\begin{document}

\title{Holographic Schwinger effect in a rotating strongly coupled medium}

\author{Yi-ze Cai}
\affiliation{School of Mathematics and Physics, China University
of Geosciences, Wuhan 430074, China}

\author{Rui-ping Jing}
\email{jingruiping@cug.edu.cn} \affiliation{School of Mathematics
and Physics, China University of Geosciences, Wuhan 430074, China}

\author{Zi-qiang Zhang}
\email{zhangzq@cug.edu.cn} \affiliation{School of Mathematics and
Physics, China University of Geosciences, Wuhan 430074, China}

\begin{abstract}
We perform the potential analysis for the holographic Schwinger
effect in a rotating deformed AdS black-hole background. We
calculate the total potential of a quark-antiquark ($Q\bar{Q}$)
pair in an external electric field and evaluate the critical
electric field from Dirac-Born-Infeld (DBI) action. It is shown
that the inclusion of angular velocity decreases the potential
barrier thus enhancing the Schwinger effect, opposite to the
effect of the confining scale. Moreover, by increasing angular
velocity decreases the critical electric field above which the
pairs are produced freely without any suppression. Furthermore, we
conclude that producing $Q\bar{Q}$ pairs would be easier in
rotating medium.
\end{abstract}

\maketitle
\section{Introduction}
One of the interesting phenomenons in quantum electrodynamics
(QED) is the pair production in a strong electric field, which is
known as the Schwinger effect \cite{JS}. The production rate of
charged particles like electron and positron was firstly computed
by Schwinger for weak-coupling and weak-field \cite{JS}
\begin{equation}
\Gamma\sim exp\Big({\frac{-\pi m^2}{eE}}\Big),
\end{equation}
where $E$, $e$ and $m$ are the external electric field, elementary
electric charge and electron mass, respectively. One finds that
there is no critical field in this scenario. Subsequently, Affleck
et.al extended the calculation of $\Gamma$ to the case for
arbitrary-coupling and weak-field \cite{IK}
\begin{equation}
\Gamma\sim exp\Big({\frac{-\pi m^2}{eE}+\frac{e^2}{4}}\Big),
\end{equation}
one finds that there is a critical field at
$E_c=(4\pi/e^3)m^2\simeq137m^2/e$, but this critical value does
not satisfy the weak-field condition, i.e., $eE\ll m^2$.
Therefore, it seems that one can not get $E_c$ under the
weak-field condition. One step further, to verify the existence of
$E_c$, one needs to work beyond the weak-field condition.

In fact, Schwinger effect is not restricted to QED, but a
universal aspect of quantum field theories (QFTs) coupled to a
U(1) gauge field. However, studying this effect in a QCD-like or
confining theory using QFTs is difficult, because the (original)
Schwinger effect is non-perturbative. Fortunately, the AdS/CFT
correspondence
\cite{Maldacena:1997re,Gubser:1998bc,MadalcenaReview} provides yet
another way. In this approach, Semenoff and Zarembo pioneered the
holographic Schwinger effect and found \cite{GW}
\begin{equation}
\Gamma\sim
exp\Big[-\frac{\sqrt{\lambda}}{2}\Big(\sqrt{\frac{E_c}{E}}-\sqrt{\frac{E}{E_c}}\Big)^2\Big],
\qquad E_c=\frac{2\pi m^2}{\sqrt{\lambda}},\label{gama}
\end{equation}
where $\lambda$ is the 't Hooft coupling constant and $m$ denotes
the mass of the fundamental scalar fields in the W-boson
supermultiplet, e.g., W-bosons or quarks. Interestingly, the
critical value agrees with DBI result \cite{YS0}. Since then,
there has been a growing interest in studying holographic
Schwinger effect in this direction
\cite{YS,YS1,SCH,KB,MG,ZQ0,ZQ,ZQ1,LS,WF,ZR,KHA,KHA1,XW,KG} (for a
recent review see \cite{DK}).

Here we extend the study of the (holographic) Schwinger effect in
rotating medium using potential analysis. It was argued that
\cite{nat,zt,fb,xg,lg} the quark gluon plasma (QGP) produced in
(typical) noncentral heavy-ion collisions may carry a nonzero
angular momentum (related to colliding nuclei) on the order of
$10^4$-$10^5\hbar$ with local angular velocity in the range of
0.01-0.1 GeV. The major part of angular momentum will be taken
away by the spectator nucleons, but some amount angular momentum
might remain in the medium \cite{mib,dek,yj}. Certainly, there is
little hope of getting a significant correction due to the angular
velocity in current experiments, but it may be observed in the
near future. Actually, AdS/CFT can be as insightful in this issue
and various observables or quantities have already been studied.
Such as drag force \cite{iy,iy1,ana}, jet quenching parameter
\cite{js,bmc}, energy loss \cite{kb,mat,df}, phase transition
\cite{xc} and free energy \cite{jz}. Inspired by this, in this
paper we investigate the effect of angular velocity on the
Schwinger effect in a deformed AdS black-hole background. In
particular, we want to know how angular velocity affects the
production rate in this case. Also, this work could be regarded as
an extension of \cite{YS} to the case with confining scale and
angular velocity.

The paper is structured as follows. In the next section, we
briefly review the rotating background considered in this work. In
section 3, we perform the potential analysis for the Schwinger
effect in this background and analyze how angular velocity affects
the production rate. Moreover, we determine the critical field
from DBI action. Finally, the results and directions for future
research are discussed in section 4.

\section{Setup}
Holographic QCD models like hard wall \cite{H1,H2}, soft wall
\cite{AKE,PC} and improved holographic QCD
\cite{JP,AST,DL,DL1,SH,SH1} have achieved considerable success in
describing various aspects of hadron physics. Here we adopt a type
of soft wall model \cite{PC}
\begin{equation}
ds^2=\frac{r^2h(r)}{R^2}[-f(r)dt^2+dx^2+dy^2+dz^2]+\frac{R^2h(r)}{r^2f(r)}dr^2,\label{metric}
\end{equation}
with
\begin{equation}
f(r)=1-\frac{r_t^4}{r^4}, \qquad h(r)=e^{c^2R^4/r^2},
\end{equation}
where $R$ is the radius of AdS (hereafter we set $R=1$ for
convenience). $r$ is the radial coordinate describing the 5th
dimension. The horizon is $r=r_t$, defined by $f(r_t)=0$. The
boundary is $r=\infty$. Moreover, $h(r)$ refers to the warp
factor, determining the characteristics of the soft wall model.
$c$ stands for the deformation parameter (or confining scale),
determining the deviation from conformality.

According to \cite{mb,ce,am}, one can extend (\ref{metric}) to a
rotating case by operating a Lorentz boost in the $t-\phi$ plane
\begin{equation}
t\rightarrow \gamma(t+\omega l^2\phi), \qquad \phi\rightarrow
\gamma(\phi+\omega l^2 t),\label{bo}
\end{equation}
with
\begin{equation}
\gamma=\frac{1}{\sqrt{1-\omega^2l^2}},
\end{equation}
where $\phi$ is the angular coordinate describing the rotation.
$\omega$ represents the angular velocity. $l$ denotes the radius
of the rotating axis. Here we will focus on the qualitative
results, so we simply take $l=1 GeV^{-1}$, similar to \cite{xc}.

Given that, the corresponding transformation of (\ref{metric})
becomes
\begin{equation}
ds^2=-m(r)dt^2+r^2h(r)(dx^2+dy^2)+\frac{h(r)}{r^2f(r)}dr^2+n(r)(d\phi+p(r)dt)^2,\label{metric2}
\end{equation}
with
\begin{equation}
m(r)=\frac{h(r)f(r)r^2(1-\omega^2)}{1-f(r)\omega^2},\qquad
n(r)=h(r)r^2\gamma^2(1-f(r)\omega^2), \qquad
p(r)=\frac{\omega(1-f(r))}{1-f(r)\omega^2}.
\end{equation}

The Hawking temperature of the black hole reads
\begin{equation}
T=\frac{r_t}{\pi}\sqrt{1-\omega^2}. \label{T}
\end{equation}

Notice that for $\omega=0$, (\ref{metric2}) returns to
(\ref{metric}), while for $\omega=c=0$, it recovers to AdS black
hole.

\section{Potential analysis in (holographic) Schwinger effect}
We now proceed to study the behavior of the Schwinger effect for
the background (\ref{metric2}) following \cite{YS}. As the
transformation (\ref{bo}) is a boost in the $t-\phi$ plane, we
intend to consider the $Q\bar{Q}$ pair located at $x-y$ plane£¬
e.g., the $Q\bar{Q}$ axis is supposed to be aligned in the $x$
direction,
\begin{equation}
t=\tau, \qquad x=\sigma, \qquad y=0,\qquad \phi=0, \qquad
r=r(\sigma). \label{par}
\end{equation}

The Nambu-Goto action is given by
\begin{equation}
S=T_F\int d\tau d\sigma\mathcal L=T_F\int d\tau d\sigma\sqrt{g},
\qquad  T_F=\frac{1}{2\pi\alpha^\prime},\label{S}
\end{equation}
where $\alpha^\prime$ is related to $\lambda$ via
$\frac{R^2}{\alpha^\prime}=\sqrt{\lambda}$. $g$ is the determinant
of the induced metric
\begin{equation}
g_{\alpha\beta}=g_{\mu\nu}\frac{\partial
X^\mu}{\partial\sigma^\alpha} \frac{\partial
X^\nu}{\partial\sigma^\beta},
\end{equation}
with $g_{\mu\nu}$ and $X^\mu$ the metric and target space
coordinate, respectively.

Under the ansatz (\ref{par}), the induced metric can be written as
\begin{equation}
g_{00}=-m(r)+n(r)p^{2},\qquad g_{01}=g_{10}=0,\qquad
g_{11}=\frac{h(r)}{r^{2}f(r)}\dot{r}^{2}+r^{2}h(r)
\end{equation}
which yields
\begin{equation}
\mathcal{L}=\sqrt{A(r)+B(r)\dot{r}^{2}}, \label{L}
\end{equation}
with
\begin{equation}
A(r)=[m(r)-n(r)p^{2}(r)]r^{2}h(r),\qquad
B(r)=\frac{[m(r)-n(r)p^{2}(r)]h(r)}{r^{2}f(r)},
\end{equation}
where $\dot{r}=\frac{dr}{d\sigma}$.

Note that $\mathcal L$ does not depend on $\sigma$ explicitly, so
the Hamiltonian is conserved,
\begin{equation}
\mathcal L-\frac{\partial\mathcal
L}{\partial\dot{r}}\dot{r}=Constant.
\end{equation}

By imposing the boundary condition
\begin{equation}
\frac{dr}{d\sigma}=0,\qquad  r=r_c\qquad (r_t<r_c<r_0)\label{con},
\end{equation}
one finds
\begin{equation}
\frac{dr}{d\sigma}=\sqrt{\frac{A^2(r)-A(r)A(r_c)}{A(r_c)B(r)}}\label{dotr},
\end{equation}
with $A(r_c)=A(r)|_{r=r_c}$, here $r=r_0$ is an intermediate
position in the bulk and this can yield a finite mass \cite{GW}.
The configuration of the string world-sheet is shown in fig.1.
\begin{figure}
\centering
\includegraphics[width=10cm]{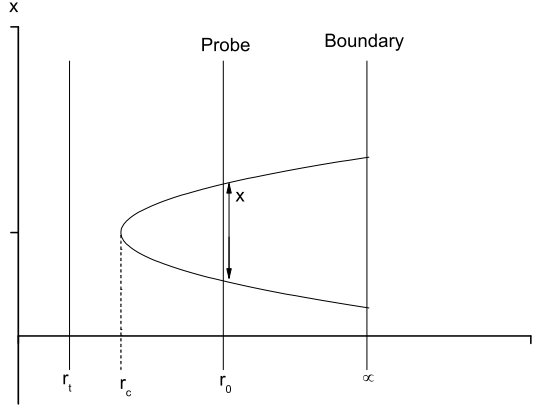}
\caption{String configuration.}
\end{figure}

From (\ref{con}) and (\ref{dotr}), the inter-distance between
$Q\bar{Q}$ is obtained
\begin{equation}
x=2\int_{r_c}^{r_0}dr\sqrt{\frac{A(r_c)B(r)}{A^2(r)-A(r)A(r_c)}}\label{xx}.
\end{equation}

Substituting (\ref{L}) and (\ref{dotr}) into (\ref{S}), the sum of
Coulomb potential and static energy is obtained
\begin{equation}
V_{CP+E}=2T_F\int_{r_c}^{r_0}dr\sqrt{\frac{A(r)B(r)}{A(r)-A(r_c)}}.\label{en}
\end{equation}

Next, we calculate the critical field. The DBI action is given by
\begin{equation}
S_{DBI}=-T_{D3}\int
d^4x\sqrt{-det(G_{\mu\nu}+\mathcal{F}_{\mu\nu})}\label{dbi},
\end{equation}
with
\begin{equation}
T_{D3}=\frac{1}{g_s(2\pi)^3\alpha^{\prime^2}}, \qquad
\mathcal{F}_{\mu\nu}=2\pi\alpha^\prime F_{\mu\nu},
\end{equation}
where $T_{D3}$ refers to the D3-brane tension.

The induced metric reads
\begin{equation}
G_{00}=-m(r)+n(r)p^2(r), \qquad G_{11}= G_{22}=r^2h(r),\qquad
G_{33}=n(r),\qquad G_{04}= G_{40}=n(r)p(r).
\end{equation}

If the electric field is turned on along the $x$ direction
\cite{YS}, then
\begin{equation}
G_{\mu\nu}+\mathcal{F}_{\mu\nu}=\left(
\begin{array}{cccc}
-m(r)+n(r)p^2(r) & 2\pi\alpha^\prime E & 0 & n(r)p(r)\\
 -2\pi\alpha^\prime E & r^2h(r) & 0 & 0 \\
 0 & 0 & r^2h(r) & 0\\
n(r)p(r) & 0 & 0 & n(r)
\end{array}
\right),
\end{equation}
yielding
\begin{equation}
det(G_{\mu\nu}+\mathcal{F}_{\mu\nu})=r^2h(r)n(r)[(2\pi\alpha^\prime)^2E^2-r^2h(r)m(r)].\label{det}
\end{equation}

Plugging (\ref{det}) into (\ref{dbi}) and making the probe
D3-brane located at $r=r_0$, one gets
\begin{equation}
S_{DBI}=-T_{D3}r_0\sqrt{h_0n_0}\int d^4x
\sqrt{r_0^2h_0m_0-(2\pi\alpha^\prime)^2E^2}\label{dbi1},
\end{equation}
with $h_0=h(r)|_{r=r_0}$, $m_0=m(r)|_{r=r_0}$, etc.

To avoid the action (\ref{dbi1}) being ill-defined, one has
\begin{equation}
r_0^2h_0m_0-(2\pi\alpha^\prime)^2E^2\geq0,\label{ec}
\end{equation}
results in
\begin{equation}
E\leq\frac{r_0}{2\pi\alpha^\prime}\sqrt{h_0m_0}=T_Fr_0\sqrt{h_0m_0}.
\end{equation}

As a result, the critical field is
\begin{equation}
E_c=T_Fr_0\sqrt{h_0m_0},\label{ec1}
\end{equation}
one can check that $E_c$ depends on $T$, $c$ and $\omega$.

Also, it is instructive to give the mass of the fundamental matter
\begin{equation}
m=T_F\int_{r_h}^{r_0}\sqrt{-det g_{ab}}=T_F(r_0-r_t),
\end{equation}
on writing $m=m_0+\Delta m$, where $m_0=T_Fr_0$ is the mass in the
system without $T$ and $\omega$, $\Delta m=-T_Fr_t$ depends on $T$
and $\omega$, one can rewrite $E_c$ as
\begin{eqnarray}
E_c&=&\frac{2\pi M^2}{\sqrt{\lambda}}(1-\frac{\Delta
m}{m})^2h_0\sqrt{\frac{f_0(1-\omega^2)}{1-f_0\omega^2}},
\end{eqnarray}
one can see that $E_c$ is well defined for $T$ and $\omega$ with
fixed $r_0$.

Now we are going to calculate the total potential. For
convenience, we introduce some dimensionless parameters as
\begin{equation}
\alpha\equiv\frac{E}{E_c}, \qquad y\equiv\frac{r}{r_c},\qquad
a\equiv\frac{r_c}{r_0},\qquad b\equiv\frac{r_t}{r_0}. \label{afa}
\end{equation}

Then the total potential becomes
\begin{eqnarray}
V_{tot}(x)&=&V_{CP+E}-Ex\nonumber\\&=&2ar_0T_F\int_1^{1/a}dy\sqrt{\frac{A(y)B(y)}{A(y)-A(y_c)}}\nonumber\\&-&
2ar_0T_F\alpha
r_0\sqrt{h_0m_0}\int_1^{1/a}dy\sqrt{\frac{A(y_c)B(y)}{A^2(y)-A(y)A(y_c)}},
\label{V}
\end{eqnarray}
where
\begin{eqnarray}
A(y)&=&(m(y)-n(y)p^2(y))(ar_0y)^2h(y),\nonumber\\
B(y)&=&(m(y)-n(y)p^2(y))h(y)/((ar_0y)^2f(y))\nonumber\\
A(y_c)&=&(m(y_c)-n(y_c)p^2(y_c))(ar_0)^2h(y_c),
\end{eqnarray}
with
\begin{eqnarray}
m(y)&=&\frac{h(y)f(y)(ar_0y)^2(1-\omega^2)}{1-f(y)\omega^2},\qquad
n(y)=h(y)(ar_0y)^2\gamma^2(1-f(y)\omega^2),\qquad
p(y)=\frac{\omega(1-f(y))}{1-f(y)\omega^2},\nonumber\\
m(y_c)&=&\frac{h(y_c)f(y_c)(ar_0)^2(1-\omega^2)}{1-f(y_c)\omega^2},\qquad
n(y_c)=h(y_c)(ar_0)^2\gamma^2(1-f(y_c)\omega^2),\qquad
p(y_c)=\frac{\omega(1-f(y_c))}{1-f(y_c)\omega^2},\nonumber\\
h(y)&=&e^{\frac{c^2}{(ar_0y)^2}}, \qquad
h(y_c)=e^{\frac{c^2}{(ar_0)^2}},\qquad
f(y)=1-(\frac{b}{ay})^4,\qquad f(y_c)=1-(\frac{b}{a})^4,
\end{eqnarray}
notice that by setting $c=\omega=0$ in (\ref{V}), the result of
SYM \cite{YS} will be recovered.

\begin{figure}
\centering
\includegraphics[width=8.5cm]{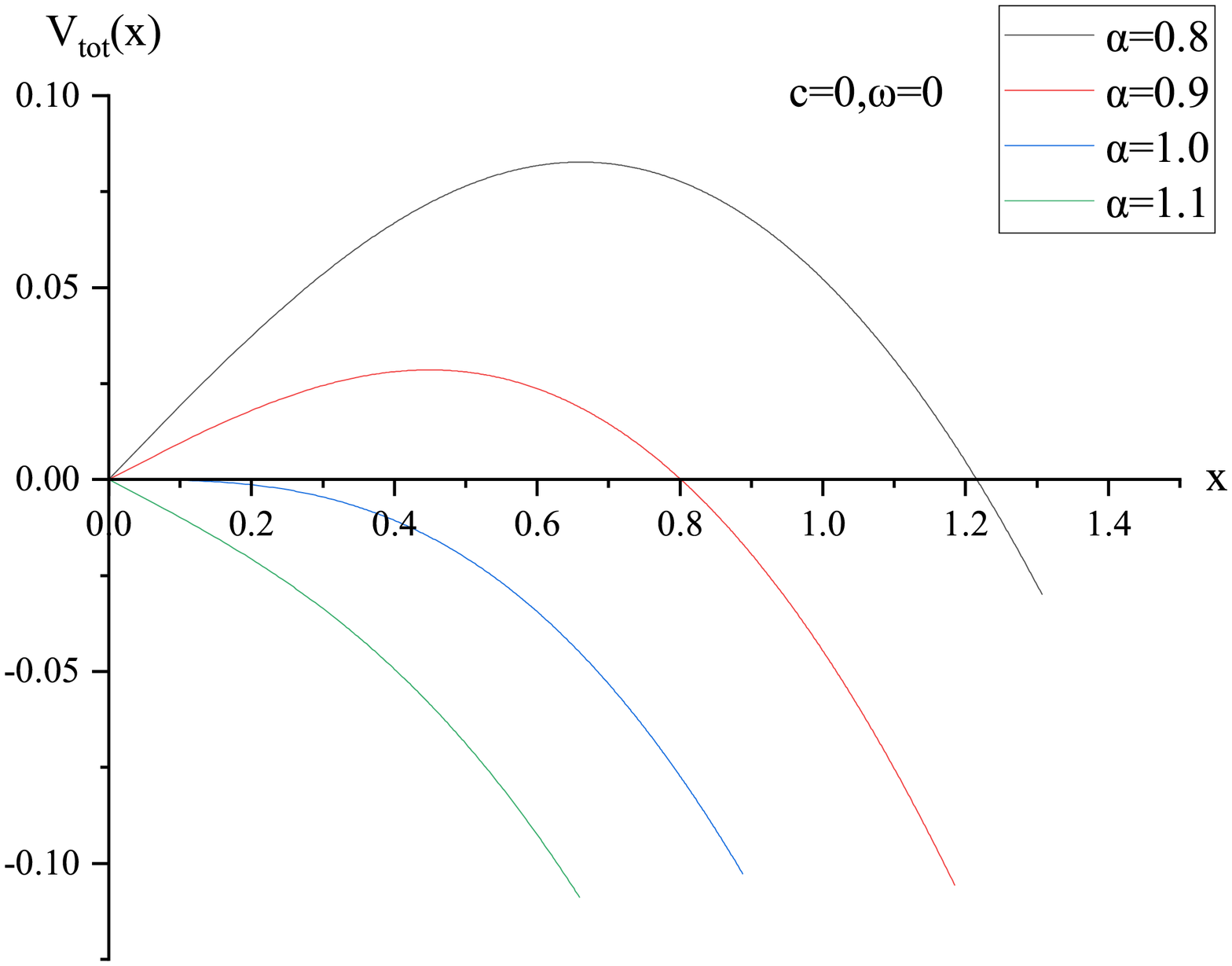}
\includegraphics[width=8.5cm]{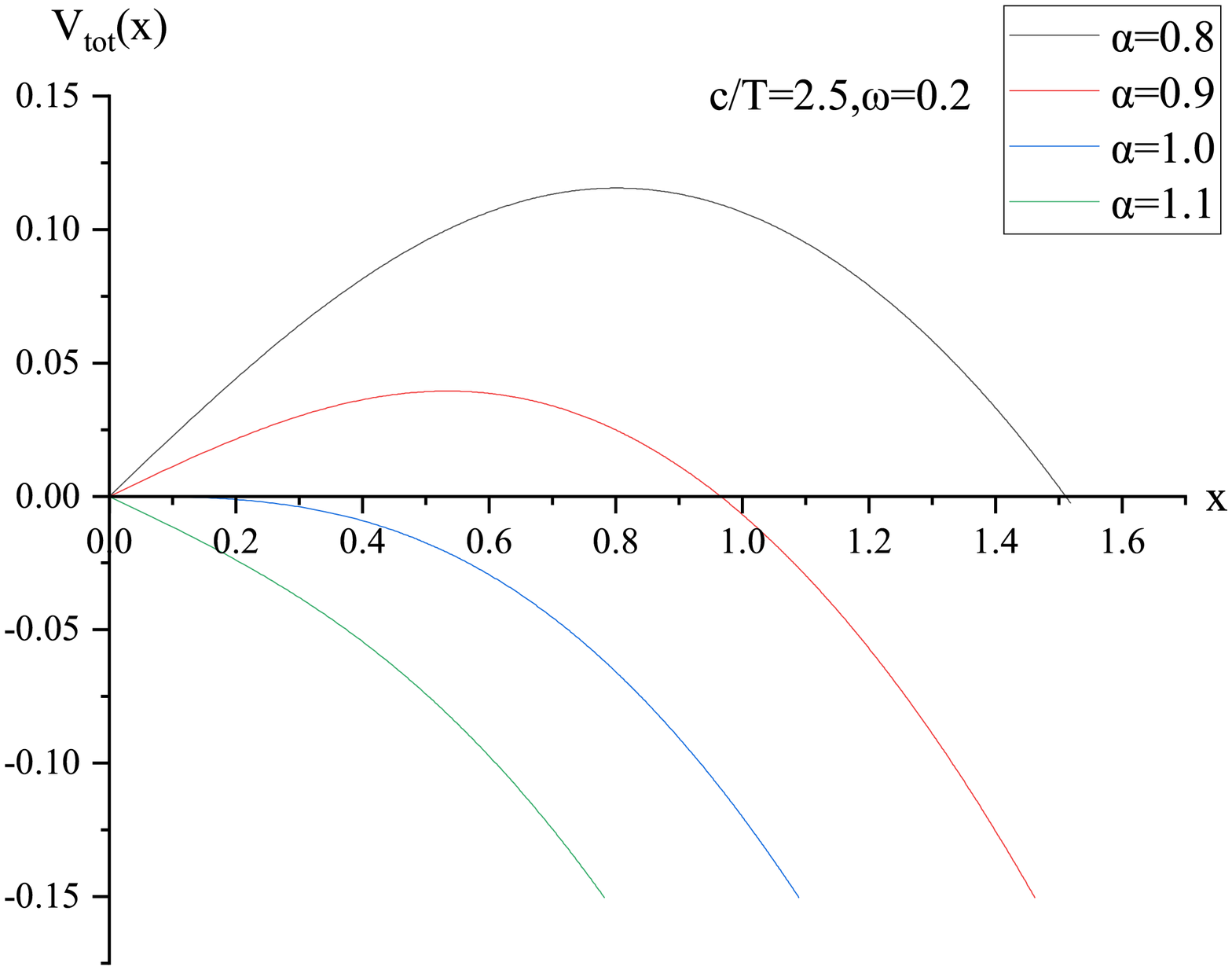}
\caption{$V_{tot}(x)$ versus $x$. Left: $c/T=0, \omega=0$. Right:
$c/T=2.5, \omega=0.2 GeV$. In both panels from top to bottom
$\alpha=0.8, 0.9, 1, 1.1$, respectively.}
\end{figure}

\begin{figure}
\centering
\includegraphics[width=8.5cm]{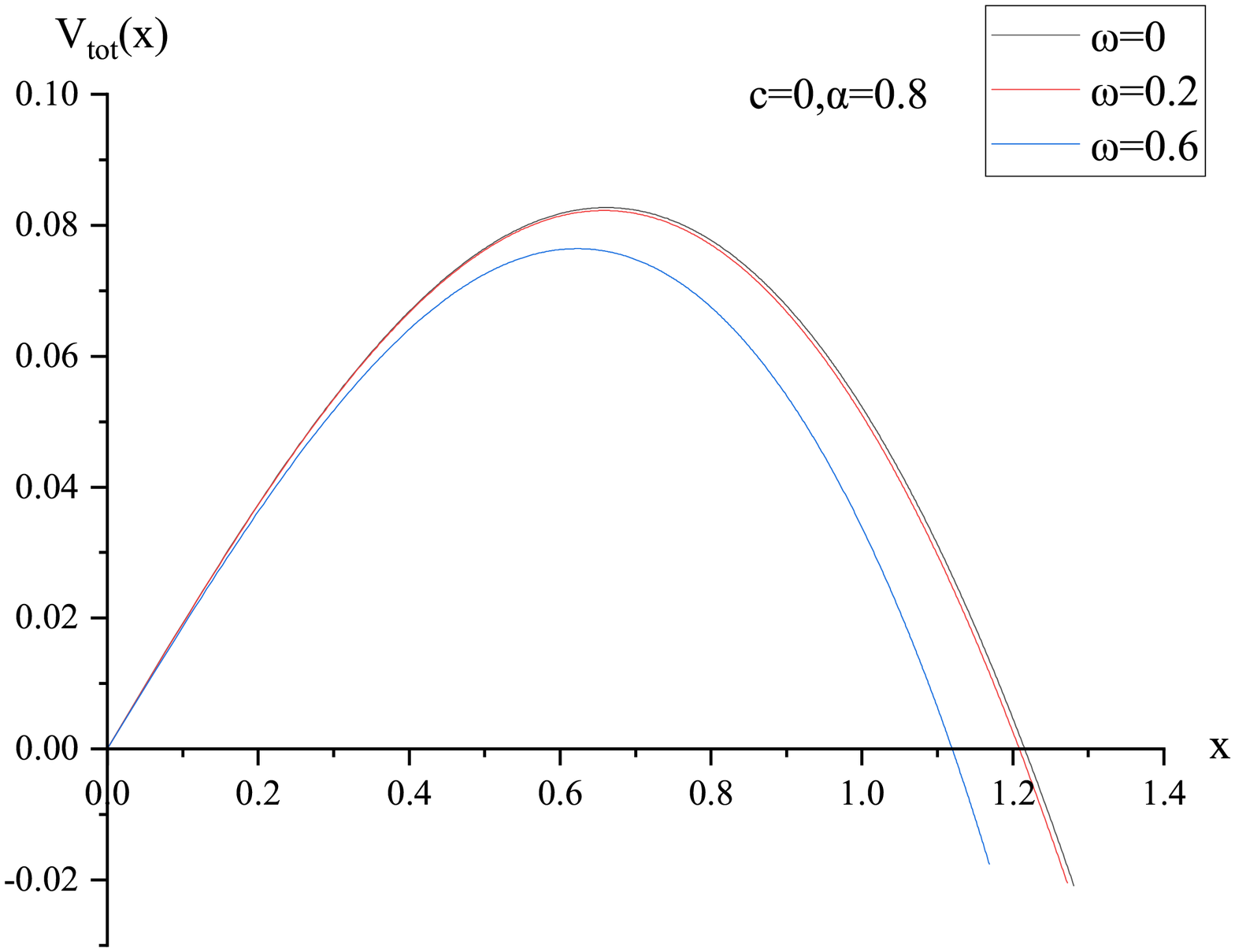}
\includegraphics[width=8.5cm]{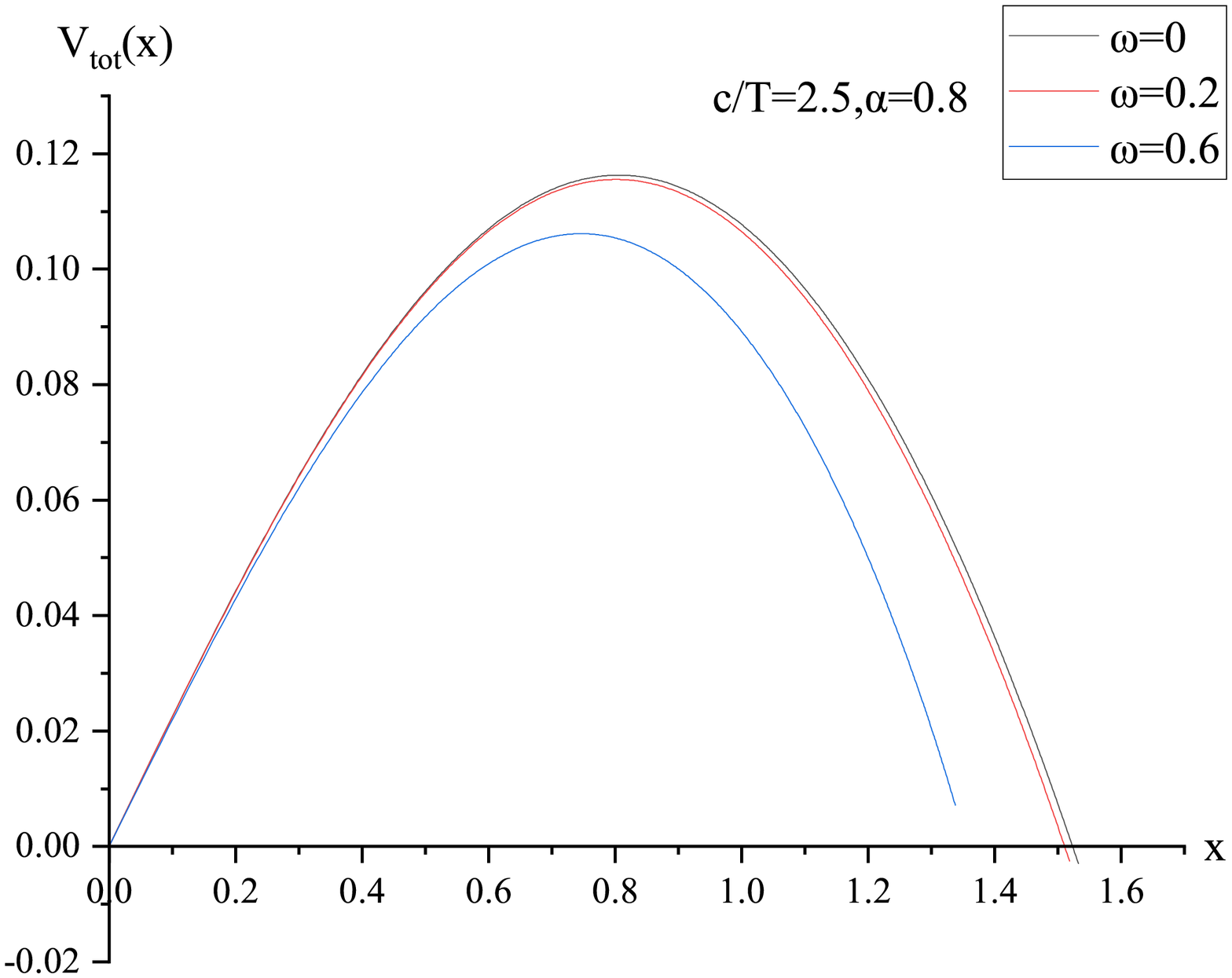}
\caption{$V_{tot}(x)$ versus $x$ with $\alpha=0.8$ and fixed $c/T$
for different values of $\omega$. In both plots from top to bottom
$\omega=0, 0.2, 0.6 GeV$, respectively.}
\end{figure}

\begin{figure}
\centering
\includegraphics[width=11cm]{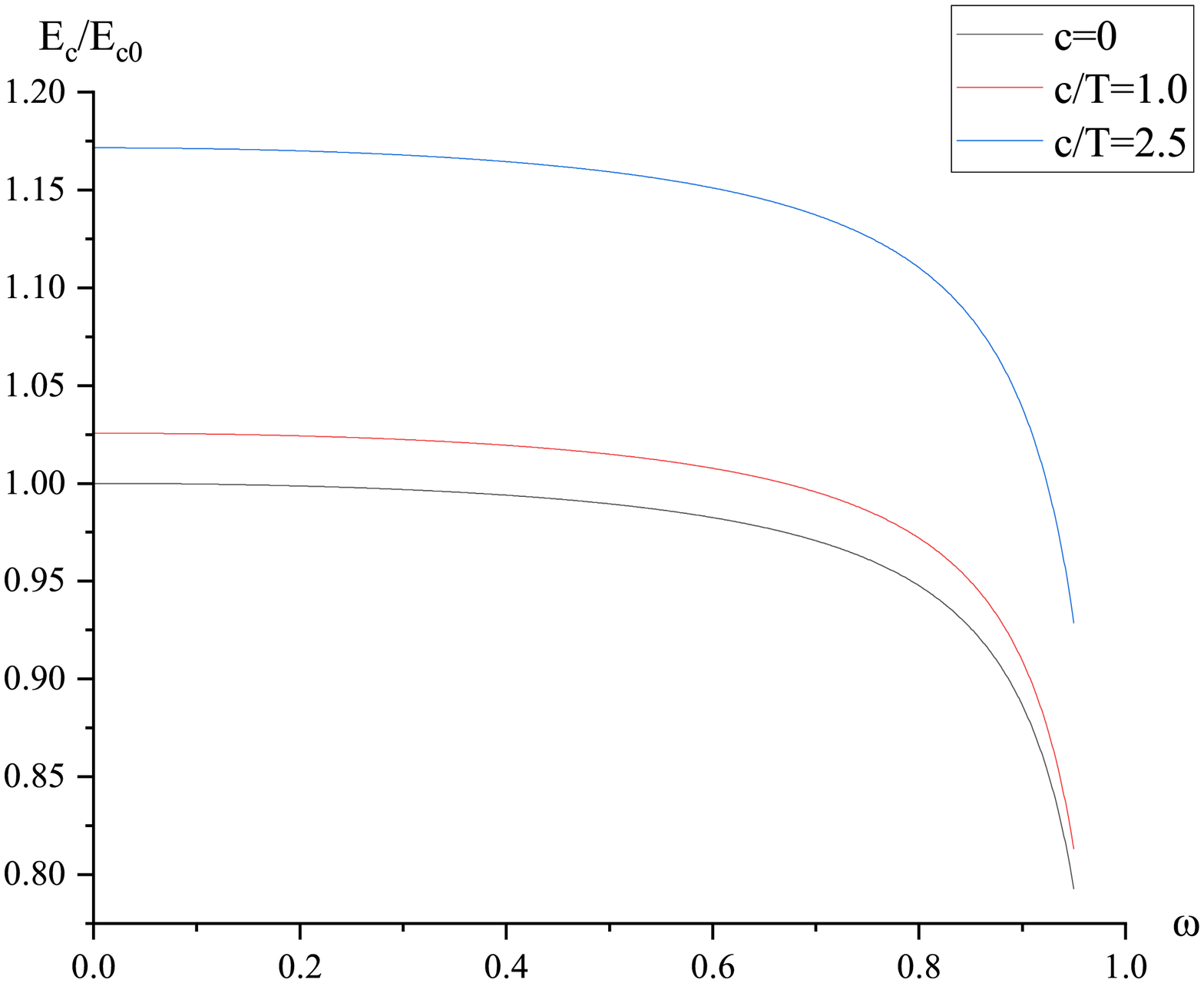}
\caption{Left: $E_c/E_{c0}$ versus $\omega$. In both plots from
top to bottom $c/T=2.5, 1, 0$, respectively.}
\end{figure}

Before proceeding, we discuss the values of some parameters.
First, for comparison purposes, we take $T_F=1$ and
$b=0.5/\sqrt{1-\omega^2}$ \cite{YS}. On the other hand, since the
range of $0\leq c/T\leq2.5$ is most relevant for a comparison with
QCD \cite{HLL}, we will choose the value of $c$ in that range.

In fig.2, we plot $V_{tot}(x)$ as a function of $x$ for various
cases, where the left panel is for $c/T=0, \omega=0$ while the
right $c/T=2.5, \omega=0.2 GeV$ (other cases with different values
of $c/T$ and $\omega$ have similar plots). From both panels, one
can see that for $\alpha<1$ (or $E<E_c$), the potential barrier
$V_{tot}(x)$ is present, then the pair production can be described
as a tunneling process. As $E$ becomes greater, $V_{tot}(x)$
decreases gradually and vanishes at $\alpha=1$ (or $E=E_c$). When
$\alpha>1$ ($E>E_c$), the pair production is catastrophic and the
vacuum becomes catastrophically unstable. These findings are in
accordance with \cite{YS}.

To investigate how angular velocity affects the Schwinger effect,
we plot $V_{tot}(x)$ against $x$ with fixed $c/T$ for different
values of $\omega$ in fig.3, where the left panel is for $c/T=0$
and the right $c/T=2.5$. In both panels from top to bottom
$\omega=0, 0.2, 0.6 GeV$, respectively. One sees that at fixed
$c/T$, the height and width of $V_{tot}(x)$ decrease with the
increase of $\omega$. It is known that the higher or the wider the
potential barrier, the harder the produced pairs escape to
infinity. Therefore, one can draw the conclusion that the
inclusion of angular velocity decrease the potential barrier thus
enhancing Schwinger effect. Namely, producing $Q\bar{Q}$ pairs
would be easier in the presence of angular velocity. Also, by
comparing the left panel with the right one, one finds that $c$
has opposite effect, i.e., increasing $c$ increases $V_{tot}(x)$
thus reducing Schwinger effect, consistent with the findings of
\cite{ZQ0}.

Moreover, to see how angular velocity modifies the critical
electric field, we plot $E_c/E_{c0}$ versus $\omega$ in fig.4,
where $E_{c0}$ represents the critical electric field of SYM. One
finds that $E_c/E_{c0}$ decreases as $\omega$ increases, implying
the inclusion of angular velocity decreases $E_c$ thus making the
tunneling process easier, in agreement with the potential
analysis. Meanwhile, $c$ has an opposite effect. In addition, one
sees that $E_c/E_{c0}$ can be smaller or larger than 1, which
means the model considered in this work could provide a wider
range of Schwinger effect in comparison to SYM.

\section{conclusion}
In this paper, we investigated the effect of angular velocity on
the Schwinger effect using potential analysis in a deformed AdS
black-hole background. We calculated the total potential of a
$Q\bar{Q}$ pair in an external electric field and determined the
critical electric field from DBI action. It is found that the
inclusion of angular velocity enhances the Schwinger effect,
opposite to the effect of confining scale $c$. Moreover, with some
chosen values of $\omega$ and $c/T$, $E_c$ can be smaller or
larger than its counterpart of SYM, indicating the model
considered here could provide theoretically a wider range of the
Schwinger effect in comparison to SYM.

Interestingly, the holographic Schwinger effect has been discussed
in a moving medium \cite{ZQ1} and the results show that the
presence of velocity increases the Schwinger effect as well. Taken
together, one may summarize that translation and rotation have the
same effect on Schwinger effect. One step further, producing
$Q\bar{Q}$ pairs would be easier in moving or rotating medium. One
likely explanation is that in moving or rotating backgrounds,
virtual pairs may not only gain energy from the external electric
field but also from the kinetic energy associated with translation
and rotation. However, the exact mechanism is not very clear and
needs more investigations.

Finally, it would be interesting to mention that the rotating QGP
may also be described by means of 5-dimensional Kerr-AdS black
hole \cite{sw}. One can study the Schwinger effect in that
rotating frame as well. We hope to report our progress in this
regard in the near future.


\end{document}